**PAPER • OPEN ACCESS**

# Signal-based classical emulation of a universal quantum computer

To cite this article: Brian R La Cour and Granville E Ott 2015 *New J. Phys.* **17** 053017

View the article online for updates and enhancements.

## Related content

- Quantum computing with photons: introduction to the circuit model, the one-way quantum computer, and the fundamental principles of photonic experiments
  Stefanie Barz

- Quantum information processing with superconducting circuits: a review
  G Wendin

- Quantum error correction for beginners
  Simon J Devitt, William J Munro and Kae Nemoto

## Recent citations

- Physical Computing: Unifying Real Number Computation to Enable Energy Efficient Computing
  Jennifer Hasler and Eric Black

- Experimental observation of classical analogy of topological entanglement entropy
  Tian Chen *et al*

- Quantum-inspired microwave signal processing for implementing unitary transforms
  Shihao Zhang *et al*

# New Journal of Physics

The open access journal at the forefront of physics

Deutsche Physikalische Gesellschaft **DPG**

**IOP** Institute of Physics

Published in partnership with: Deutsche Physikalische Gesellschaft and the Institute of Physics

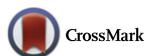

OPEN ACCESS

PAPER

## Signal-based classical emulation of a universal quantum computer

Brian R La Cour and Granville E Ott

Applied Research Laboratories, The University of Texas at Austin, PO Box 8029, Austin, TX 78713-8029, USA

E-mail: blacour@arlut.utexas.edu

## Abstract

In this paper we present a novel approach to emulating a universal quantum computer with a classical system, one that uses a signal of bounded duration and amplitude to represent an arbitrary quantum state. The signal may be of any modality (e.g., acoustic, electromagnetic, etc), but we focus our discussion here on electronic signals. Unitary gate operations are performed using analog electronic circuit devices, such as four-quadrant multipliers, operational amplifiers, and analog filters, although non-unitary operations may be performed as well. In this manner, the Hilbert space structure of the quantum state, as well as a universal set of gate operations, may be fully emulated classically. The required bandwidth scales exponentially with the number of qubits, however, thereby limiting the scalability of the approach, but the intrinsic parallelism, ease of construction, and classical robustness to decoherence may nevertheless lead to capabilities and efficiencies rivaling that of current high performance computers.

## 1. Introduction

The notion of quantum computation has ushered in a new paradigm for computational science, one that challenges our understanding of complexity theory and its relation to classical computing [1, 2]. Well-known applications such as Shor's factoring algorithm demonstrate the improved efficiency possible with quantum computers over the best known classical algorithms [3], but the intrinsic power of quantum computing was apparent even in the seminal problems of Deutsch and Jozsa [4, 5]. If, indeed, quantum computers turn out to be more efficient than their classical counterparts, it would entail nothing less than a refutation of the venerated Church–Turing thesis that one can do no better than a Turing-equivalent machine [6, 7].

The fundamental property of quantum computers that makes these efficiencies possible is the Hilbert space structure of the quantum state, which gives rise to linear superpositions of classical binary states over a complex scalar field. The quantum state is therefore one of a continuum of possible states, resembling more an analog than digital computer in this regard. Because the dimension of the Hilbert space scales exponentially with the number of quantum bits (or, qubits), a quantum computer is considered to have vastly greater capability over a digital computer with an equivalent number of classical bits.

Recognizing the importance of the Hilbert space structure of the quantum state, and the linearity of the operations upon it, as the fundamental drivers of efficiency, one may ask whether any classical systems exhibit these same mathematical structures. The appeal of such a classical system lies in its presumed relative ease of construction and robustness to decoherence. The primary drawback lies in the instrinsic limitation to the number of qubits that can practically be represented, as first noted by Feynman [8]. Nevertheless, it is interesting, from a foundational perspective, to understand whether there are classical systems capable of emulating quantum computers at all. From a more practical perspective, however, there may be great utility in building quantum emulation devices which, despite their classical nature, may nevertheless benefit from the many insights of quantum information science.

A formal analogy between classical optics and quantum computing was first recognized by Spreeuw in 1998 using a combination of laser beams and polarization modes to represent an arbitrary quantum state [9, 10]. This scheme did not provide statistical measurement outcomes, only state representations; but those states could be entangled (i.e., made non-separable) and used to implement various quantum computing algorithms. Using a similar approach, Kwiat and colleagues were able to demonstrate an implementation of Grover's search algorithm using only classical optics [11]. Although in theory such an approach is capable of representing any number of qubits, the physical resources required to do so scale exponentially, thereby limiting its applicability.

From a practical perspective, complex optical systems are notoriously difficult to build. In 2001, Ferry and coworkers proposed that a quantum computer would be equivalent to a parallel analog electronic computer [12]. They suggested that a single qubit could be represented by the in-phase and quadrature components of a modulated signal, but their scheme was not easily extensible to multiple qubits. Two years later, Laszlo Kish introduced the notion of a Hilbert space analog (HSA) computer, which uses analog electronic circuits to represent and manipulate a notional quantum state [13]. A similar concept was used to emulate a quantum computer of 8 and 16 qubits using a so-called quantum-circuit processor (QCP) [14, 15]. The device proved to be very fast, but the explicit representation of the state and gate operations did not take full advantage of the tensor product structure of the constituent qubits. Also, like Spreeuw's optical scheme, these particular classical analogies do not provide a statistical model of quantum measurement.

It is important to understand that these quantum–classical analogies do not merely *simulate* quantum computers, they *emulate* them, much as an inductive-capacitive circuit can emulate a mass-spring system. Numerous quantum computer simulators exist, developed as software implemented on standard digital computers. That one can produce such simulators is trivial—they are merely a numerical representation of matrix operations. It is perhaps rather more suprising that one can emulate classically the Hilbert space structure and time evolutionary behavior of a quantum system.

In this paper we present a new approach to emulating a universal quantum computer with a classical system, one that uses a signal of bounded duration and amplitude to represent an arbitrary quantum state. The signal may be of any modality (e.g., acoustic, electromagnetic, etc), but we focus our discussion here on electronic signals. Like Ferry *et al* [12], we use quadrature modulation to represent a single qubit, but our approach easily generalizes to multiple qubits. Unitary gate operations are performed using analog electronic circuit devices, such as four-quadrant multipliers, operational amplifiers, and analog filters, although non-unitary operations may be performed as well. Unlike the HSA and QCP approaches, which operate explicitly on the quantum state components, we perform these gate operations by decomposing the quantum state into pairs of subspace projection signals, thereby avoiding an otherwise cumbersome spectral decomposition and resynthesis process for each gate operation. Finally, using a hidden-variable model of quantum measurement, these same projection operations may be used to realize statistical measurement gates [16]. In this manner, the Hilbert space structure of the quantum state, as well as a universal set of gate operations, may be fully emulated classically.

The organization of the paper is as follows. In section 2 we describe the basic representation of a multi-qubit state in terms of a nested sequence of amplitude-modulated tonals and show that this reproduces the Hilbert space structure of a tensor product of qubits. In section 3 we outline the construction of projection operators used to address individual qubits and how they may be used to perform one- and two-qubit gate operations. This approach is then applied to the construction of unitary and measurement gates in sections 4 and 5, respectively. Sections 6 and 7 consider applications illustrating quantum teleportation and quantum parallelism. We turn to some practical considerations regarding noise and bandwidth limitations in section 8 and provide a summary of our findings and conclusions in section 9.

## 2. Quantum state representation

In this section we develop a physical representation of the Hilbert space structure of a quantum computer in terms of classical signals. In the course of this development, we will identify the necessary mathematical structures and match them with particular physical entities or operational procedures. We therefore begin with a specification of the relevant mathematical entities.

The state of an $n$-qubit quantum computer may be represented by a element $|\psi\rangle$ of a $2^n$-dimensional Hilbert space $\mathcal{H}$ taking on the particular form of a tensor product of $n$ two-dimensional Hilbert spaces, here denoted $\mathcal{H}_0, \ldots, \mathcal{H}_{n-1}$, such that $\mathcal{H} = \mathcal{H}_{n-1} \otimes \cdots \otimes \mathcal{H}_0$. (Here, and throughout this paper, we use the little endian convention of ordering the qubits from right to left.) A single element of one of the $n$ constituent Hilbert spaces constitutes a qubit. The specification of an inner product $\langle \phi | \psi \rangle$ between states $|\phi\rangle$ and $|\psi\rangle$ in $\mathcal{H}$ completes the Hilbert-space description. The norm of $|\psi\rangle$ may then be defined as $\||\psi\rangle\| := \sqrt{\langle \psi | \psi \rangle}$.

We shall denote by $|0\rangle_i$ and $|1\rangle_i$ a pair of orthonormal basis states, termed the computational basis, for $\mathcal{H}_i$ and $i \in \{0, \ldots, n-1\}$. Taking tensor products of these individual basis states, we obtain a set of $2^n$ orthonormal basis states for the product space, $\mathcal{H}$. A particular binary sequence $x_0, \ldots, x_{n-1}$ therefore corresponds to a single basis state $|x_{n-1}\rangle_{n-1} \otimes \cdots \otimes |x_0\rangle_0$. For brevity, this binary sequence may be represented by its decimal form, $x = x_0 2^0 + \cdots + x_{n-1} 2^{n-1} \in \{0, \ldots, 2^n - 1\}$, so that the corresponding basis state may be written succinctly as $|x\rangle$ or, occasionally, $|x_{n-1} \cdots x_0\rangle$. Let $\langle x|\psi\rangle = \alpha_x \in \mathbb{C}$ for a given state $|\psi\rangle \in \mathcal{H}$ and basis state $|x\rangle$. This state may then be written as

$$|\psi\rangle = \sum_{x=0}^{2^n-1} \alpha_x |x\rangle. \tag{1}$$

### 2.1. Zero-qubit states

We begin with a description of a (rather trivial) one-dimensional Hilbert space in which the 'zero-qubit' state $|\psi\rangle$ is represented by a single complex number $\alpha \in \mathbb{C}$. (Here, and in what follows, we forgo the usual constraint that $\||\psi\|\| = 1$. This normalization may be applied later, as needed.) Although quite simple, this representation will form the conceptual basis of more complex structures to follow.

Let us define the zero-qubit signal as the sum of in-phase and quadrature components with angular frequency $\omega_c > 0$ as follows [17]:

$$s(t) = a \cos(\omega_c t) - b \sin(\omega_c t), \tag{2}$$

where $a = \text{Re}[\alpha]$ and $b = \text{Im}[\alpha]$. Multiplication by the in-phase and quadrature reference signals, scaled by a factor of two, yields

$$2\cos(\omega_c t)s(t) = a\left[1 + \cos(2\omega_c t)\right] - b\sin(2\omega_c t) \tag{3a}$$

$$-2\sin(\omega_c t)s(t) = -a\sin(2\omega_c t) + b\left[1 - \cos(2\omega_c t)\right]. \tag{3b}$$

Now, applying a low-pass filter with a passband below $2\omega_c$ removes the higher frequency components and yields the in-phase and quadrature amplitudes. This may be implemented, for example, by a time average over the signal period $T \in (2\pi/\omega_c)\mathbb{N}$, since

$$\frac{1}{T}\int_0^T 2\cos(\omega_c t)s(t)\,\mathrm{d}t = a, \tag{4a}$$

$$\frac{1}{T}\int_0^T -2\sin(\omega_c t)s(t)\,\mathrm{d}t = b. \tag{4b}$$

Note that, although $s$ is a real signal, we may write it in terms of a complex signal as follows:

$$\begin{aligned}s(t) &= a\left(\frac{e^{i\omega_c t} + e^{-i\omega_c t}}{2}\right) - b\left(\frac{e^{i\omega_c t} - e^{-i\omega_c t}}{2i}\right) \\ &= \frac{1}{2}\left[(a+ib)e^{i\omega_c t} + (a-ib)e^{-i\omega_c t}\right] \\ &= \text{Re}\left[\alpha\, e^{i\omega_c t}\right]. \end{aligned} \tag{5}$$

We may therefore view $\alpha$ as a complex (DC, in this case) modulating signal with a carrier frequency of $\omega_c$, in which case we may identify the (constant) function $\psi$ given by $\psi(t) = \alpha$ as corresponding to the quantum state $|\psi\rangle$. Given a zero-qubit signal $s$, we may then obtain the corresponding zero-qubit state $\psi$ by using the above time-averaging procedure to obtain the real and imaginary parts of $\psi$. Similarly, given the state $\psi$, we may obtain the corresponding signal $s$ by modulating the in-phase and quadrature components of a carrier signal of frequency $\omega_c$.

To define an inner product, we begin by noting that a time average of $s(t)^2$ over $t \in [0, T]$ yields

$$\frac{1}{T}\int_0^T s(t)^2\,\mathrm{d}t = \frac{a^2}{2} + \frac{b^2}{2} = \frac{|\alpha|^2}{2} = \frac{\langle\psi|\psi\rangle}{2}. \tag{6}$$

More generally, for a second signal of the form $r(t) = \text{Re}[\phi(t)e^{i\omega_c t}]$, where $\phi(t) = \beta \in \mathbb{C}$, we define

$$\langle\phi|\psi\rangle = \frac{1}{T}\int_0^T \phi(t)^\star \psi(t)\,\mathrm{d}t = \beta^\star \alpha. \tag{7}$$

This completes the Hilbert space description of this simple, one-dimensional space.

It is worth noting that the zero-qubit signals considered in this section are often used as an encoding scheme in digital communications, with each value of $\alpha$ representing a different binary sequence. For example, a typical 64-QAM Ethernet protocol uses 64 different combinations of phases and amplitudes to represent a string of 6 (classical) bits. In what follows, we shall take quite a different approach to encoding information, using a nested sequence of modulating signals to represent a single $n$-qubit state in terms of its $2^n$ complex amplitudes. For this reason, we shall use the term quadrature modulated tonals (QMT) to refer to this sort of representation.

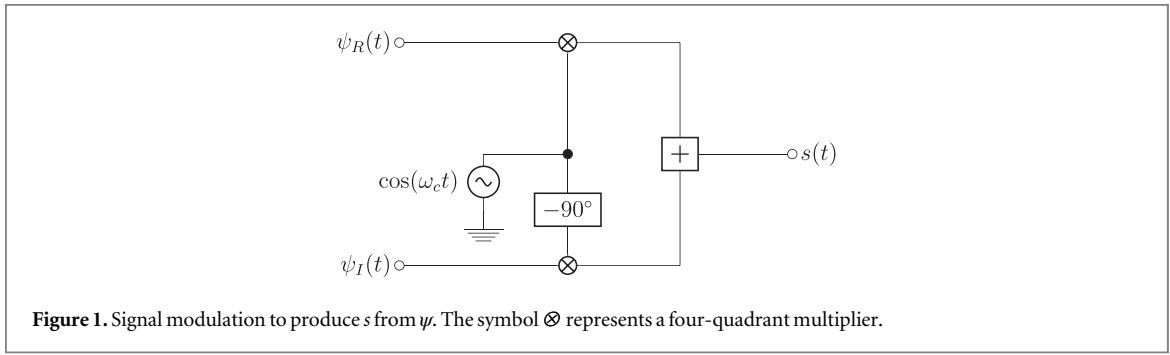

**Figure 1.** Signal modulation to produce $s$ from $\psi$. The symbol $\otimes$ represents a four-quadrant multiplier.

### 2.2. Single-qubit states

Let us turn now to representing a (simple but non-trivial) two-dimensional Hilbert space representing a single qubit. The qubit state is written as $|\psi\rangle = \alpha_0 |0\rangle + \alpha_1 |1\rangle$ in the computational basis, where $\alpha_0, \alpha_1 \in \mathbb{C}$. Much as before, let $a_0 = \mathrm{Re}[\alpha_0]$, $b_0 = \mathrm{Im}[\alpha_0]$, $a_1 = \mathrm{Re}[\alpha_1]$, $b_1 = \mathrm{Im}[\alpha_1]$.

As was done in the previous section, we wish to model the quantum state as a complex, basebanded signal modulating a carrier of frequency $\omega_c$. In other words, we seek a function $\psi : \mathbb{R} \to \mathbb{C}$ for which

$$s(t) = \mathrm{Re}\left[\psi(t) e^{i\omega_c t}\right] \tag{8}$$

represents the quantum state $|\psi\rangle$. This can be achieved by amplitude modulating the carrier (of frequency $\omega_c$) by a tonal of frequency $\omega < \omega_c$.

For the in-phase and quadrature components of the complex basebanded signal, it is convenient to use complex exponentials rather than sines and cosines. Thus, we write

$$\psi(t) = \alpha_0 e^{i\omega t} + \alpha_1 e^{-i\omega t} \tag{9}$$

and identify the functions $\phi_0^\omega$ and $\phi_1^\omega$, given by

$$\phi_0^\omega(t) := e^{i\omega t}, \tag{10a}$$

$$\phi_1^\omega(t) := e^{-i\omega t}, \tag{10b}$$

with the computational basis functions $|0\rangle$ and $|1\rangle$, respectively.

Substituting these expressions, we find that the real one-qubit signal $s$ is given by the complex one-qubit state $\psi$ as follows:

$$s(t) = \psi_R(t) \cos(\omega_c t) - \psi_I(t) \sin(\omega_c t), \tag{11}$$

where $\psi(t) = \psi_R(t) + i\psi_I(t)$ and

$$\psi_R(t) = (a_0 + a_1)\cos(\omega t) - (b_0 + b_1)\sin(\omega t), \tag{12a}$$

$$\psi_I(t) = (b_0 - b_1)\cos(\omega t) - (a_0 - a_1)\sin(\omega t). \tag{12b}$$

In this way we can see that the real and imaginary parts of $\psi$ serve as the in-phase and quadrature components modulating the carrier signal. A representative circuit schematic of this process is shown in figure 1.

The mapping from $\psi$ to $s$ can be reversed through a process of demodulation. As in the zero-qubit case, this may be accomplished by alternately multiplying $s(t)$ by the in-phase, $\cos(\omega_c t)$, and quadrature, $-\sin(\omega_c t)$, signals and then low-pass filtering. The result is

$$\frac{1}{T}\int_{t-T}^{t} 2\cos(\omega_c t') s(t')\, \mathrm{d}t' = \psi_R(t), \tag{13a}$$

$$\frac{1}{T}\int_{t-T}^{t} -2\sin(\omega_c t') s(t')\, \mathrm{d}t' = \psi_I(t). \tag{13b}$$

A representative circuit schematic of this process is shown in figure 2. Because the mapping from $\psi$ to $s$ is reversible, we may focus our discussion in what follows to the former.

To complete the Hilbert space description of this one-qubit representation, we must define an inner product. Let $\psi$ be defined as before and let $\phi$ be an arbitrary one-qubit state of the form

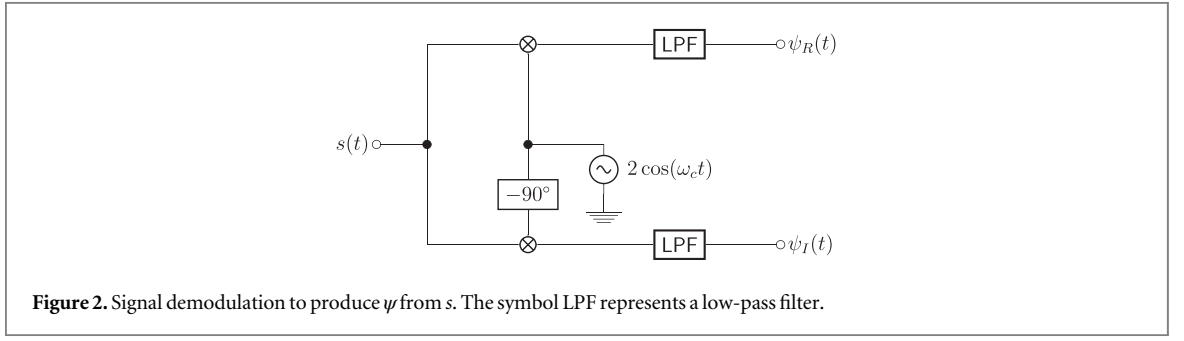

Figure 2. Signal demodulation to produce $\psi$ from $s$. The symbol LPF represents a low-pass filter.

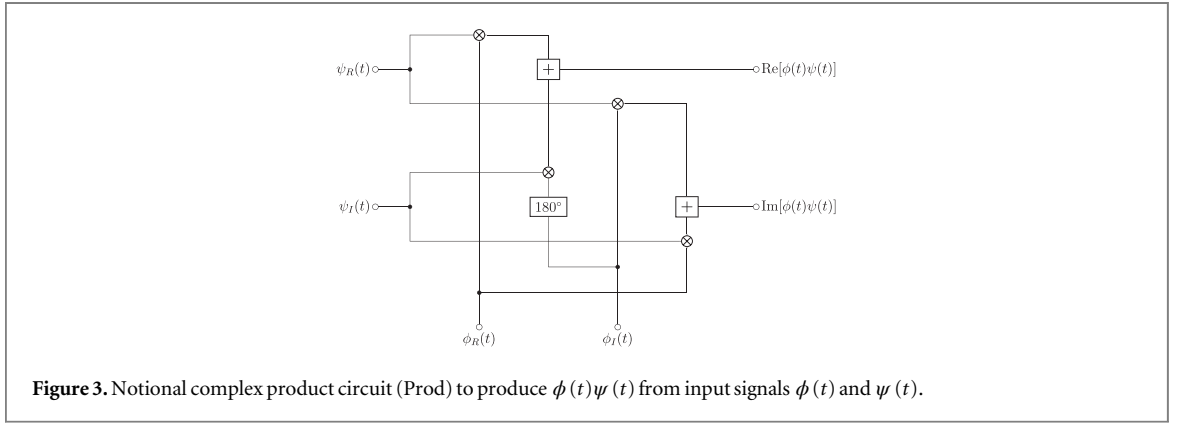

Figure 3. Notional complex product circuit (Prod) to produce $\phi(t)\psi(t)$ from input signals $\phi(t)$ and $\psi(t)$.

$$\phi(t) = \beta_0 \phi_0^\omega(t) + \beta_1 \phi_1^\omega(t) \tag{14}$$

corresponding to the quantum state $|\phi\rangle = \beta_0 |0\rangle + \beta_1 |1\rangle$. The inner product $\langle\phi|\psi\rangle$ between the two is again defined to be the time average over the period $T \in (2\pi/\omega)\mathbb{N}$:

$$\langle\phi|\psi\rangle = \frac{1}{T}\int_0^T \phi(t)^\star \psi(t)\, dt. \tag{15}$$

A representative circuit schematic implementing the complex product function is shown in figure 3. The inner product is formed by first applying a conjugate (Conj) operation to $\phi$ (i.e., inverting $\phi_I$) before multiplying it by $\psi$, then passing the resultant real and imaginary parts through low-pass filters.

Performing this integration, one readily verifies that

$$\langle\phi|\psi\rangle = \beta_0^\star \alpha_0 + \beta_1^\star \alpha_1, \tag{16}$$

so the definition produces the correct result. This definition also allows us to identify $\langle\phi|$, the dual of $|\phi\rangle$, as the functional which maps a complex basebanded signal $\psi$ to the corresponding inner product $\langle\phi|\psi\rangle$.

From the definition of the inner product we can also see that the one-qubit computational basis states $\phi_0^\omega$ and $\phi_1^\omega$ are orthonormal. This allows us to extract the components of the state by performing a time average, since

$$\langle\phi_0^\omega|\psi\rangle = \langle 0|\psi\rangle = \alpha_0, \tag{17a}$$

$$\langle\phi_1^\omega|\psi\rangle = \langle 1|\psi\rangle = \alpha_1. \tag{17b}$$

This analysis procedure may be complemented by one of synthesis wherein the state $\psi$ may be constructed from the components $\alpha_0$ and $\alpha_1$ by multiplying each by the corresponding basis state and summing the two resulting signals.

### 2.3. Two-qubit states

We are now ready for the (highly non-trivial) model of a four-dimensional Hilbert space representing a two-qubit state. The general qubit state is

$$|\psi\rangle = \alpha_{00}|00\rangle + \alpha_{01}|01\rangle + \alpha_{10}|10\rangle + \alpha_{11}|11\rangle. \tag{18}$$

We define $\alpha_{00} = a_{00} + ib_{00}$, ..., $\alpha_{11} = a_{11} + ib_{11}$, as before, and represent the two-qubit state by a complex basebanded signal of the form

$$\psi(t) = e^{i\omega_A t}\left[\alpha_{00}e^{i\omega_B t} + \alpha_{01}e^{-i\omega_B t}\right] + e^{-i\omega_A t}\left[\alpha_{10}e^{i\omega_B t} + \alpha_{11}e^{-i\omega_B t}\right], \quad (19)$$

where $\omega_A > \omega_B$ by convention. The corresponding real signal is, as before, $s(t) = \text{Re}[\psi(t)e^{i\omega_c t}]$, where we assume that the carrier frequency is sufficiently large so that $\omega_A + \omega_B < \omega_c$.

Much as before, we define $\phi_0^{\omega_A}(t) = e^{i\omega_A t}$ and $\phi_1^{\omega_A}(t) = e^{-i\omega_A t}$ and identify the basis functions $\phi_0^{\omega_A}$ and $\phi_1^{\omega_A}$ with the qubit states $|0\rangle_A$ and $|1\rangle_A$, respectively. Similarly, the basis functions $\phi_0^{\omega_B}$ and $\phi_1^{\omega_B}$ are identified with the computational basis states $|0\rangle_B$ and $|1\rangle_B$, respectively. Finally, the function product $\phi_0^{\omega_A} \cdot \phi_1^{\omega_B} = \phi_1^{\omega_B} \cdot \phi_0^{\omega_A}$, say, will be identified with the tensor product $|0\rangle_A \otimes |1\rangle_B = |1\rangle_B \otimes |0\rangle_A = |01\rangle$. Note that, unlike the Kronecker product between two matrices, the function product between qubits is commutative. Thus, the complex basebanded signal $\psi$ may be identified with the two-qubit quantum state $|\psi\rangle$.

Note that the signal $\psi$ consists of four distinct frequencies: (1) $\omega_A + \omega_B > 0$, corresponding to $\alpha_{00}$, (2) $\omega_A - \omega_B > 0$, corresponding to $\alpha_{01}$, (3) $-\omega_A + \omega_B < 0$, corresponding to $\alpha_{10}$, and (4) $-\omega_A - \omega_B < 0$, corresponding to $\alpha_{11}$. The modulated signal given by $s(t) = \text{Re}[\psi(t)e^{i\omega_c t}]$ consists of these four frequencies shifted to the right by $+\omega_c$ and to the left by $-\omega_c$. Thus, $s$ consists of eight frequencies in all, ranging from $-\omega_c - \omega_A - \omega_B$ to $\omega_c + \omega_A + \omega_B$, four of which are positive and four of which are negative, as befits a real signal.

Given the set of four complex numbers $\alpha_{00}, \ldots, \alpha_{11}$, we may produce the two-qubit signal $s$ by first constructing the real and imaginary parts, $\psi_R$ and $\psi_I$ respectively, of $\psi$. As before, $s$ may be produced by using $\psi_R$ as the in-phase signal and $\psi_I$ as the quadrature signal for a carrier of frequency $\omega_c$, to wit:

$$s(t) = \psi_R(t)\cos(\omega_c t) - \psi_I(t)\sin(\omega_c t). \quad (20)$$

For example, suppose $|\psi\rangle$ is the entangled (i.e., non-separable) Bell state $|\psi\rangle = |01\rangle + |10\rangle$. (Note, again, that we have omitted trivial normalization factors.) In this case, $\psi(t) = e^{i(\omega_A - \omega_B)t} + e^{-i(\omega_A - \omega_B)t} = 2\cos[(\omega_A - \omega_B)t]$, so $s(t) = 2\cos[(\omega_A - \omega_B)t]\cos(\omega_c t)$. By contrast, the separable state $|\psi\rangle = |00\rangle + |01\rangle$ would be represented by $\psi(t) = 2e^{i\omega_A t}\cos(\omega_B t)$, so $s(t) = 2\cos(\omega_B t)\cos[(\omega_A + \omega_c)t]$. So, in this representation, a separable state is such that one qubit modulates another; whereas, an entangled state results from, in this case, a difference of frequencies.

## 2.4. General $n$-qubit states

The generalization to $n$ qubits is now straightforward. Let $x \in \{0, 1, \ldots, 2^n - 1\}$ and define $\phi_x$ as the function product such that

$$\phi_x(t) = \phi_{x_{n-1}}^{\omega_{n-1}}(t) \cdots \phi_{x_1}^{\omega_1}(t) \cdot \phi_{x_0}^{\omega_0}(t), \quad (21)$$

where $\phi_0^{\omega_k}(t) = e^{i\omega_k t}$, $\phi_1^{\omega_k}(t) = e^{-i\omega_k t}$, and $x_0, \ldots, x_{n-1}$ are the binary digits of $x$.

By convention, we assume that $0 < \omega_0 < \cdots < \omega_{n-1}$. The $n$-qubit signal can now be written in the form $s(t) = \text{Re}[\psi(t)e^{i\omega_c t}]$, where $\omega_{n-1} + \cdots + \omega_0 < \omega_c$ and

$$\psi(t) = \sum_{x=0}^{2^n - 1} \alpha_x \phi_x(t). \quad (22)$$

For two such signals $\phi$ and $\psi$, the inner product is defined, as before, to be

$$\langle \phi | \psi \rangle = \frac{1}{T}\int_0^T \phi(t)^*\psi(t)\,dt, \quad (23)$$

where the time average is over a multiple of the period, $T \in (2\pi/\omega_0)\mathbb{N}$, of the lowest frequency qubit. This completes the Hilbert space description.

To represent $s$, we need $n + 1$ frequencies, i.e., one for each of the $n$ qubits plus a carrier frequency $\omega_c$. The product of these $n + 1$ frequencies in $s$ will have spectral components at the sums and differences of these frequencies. This can be achieved most easily by taking $\omega_i = 2^i\Delta\omega$ and $\omega_c = \omega_b + 2^n\Delta\omega$, where $\Delta\omega > 0$ and $\omega_b \geq 0$ is some baseband offset. We shall refer to this as the *octave spacing scheme*. The positive frequencies of $s$ will therefore range from $\omega_{\min} = \omega_c - (\omega_{n-1} + \cdots + \omega_0) = \omega_b + 2^n\Delta\omega - (2^n - 1)\Delta\omega = \omega_b + \Delta\omega$ to $\omega_{\max} = \omega_c + (\omega_{n-1} + \cdots + \omega_0) = \omega_b + (2^{n+1} - 1)\Delta\omega$, in increments of $2\Delta\omega$, so there will be $2^n$ different (positive) frequencies in all. The ordering of the frequencies is such that the complex coefficient $\alpha_x$ is encoded in the frequencies $\pm\omega_c + \Omega_x$, where $\Omega_x = (2^n - 1 - 2x)\Delta\omega$ for $x \in \{0, \ldots, 2^n - 1\}$.

In the octave spacing scheme, each qubit corresponds to one of $n$ frequencies, while each basis state corresponds to one of $2^n$ frequencies. We shall refer to $\omega_i$ as the *qubit frequency* for qubit $i$ and refer to $\Omega_x$ as the *basis frequency* for basis state $|x\rangle$. To synthesize a quantum state $\psi$, then, one may explicitly define each of the $2^n$ complex components, multiply this by the corresponding basis states, and sum the resulting products. Likewise, to analyze a given quantum state $\psi$, one may multiply by each of the $2^n$ basis states, compute the inner product by time averaging, and thereby obtain the corresponding complex component. Using such a procedure, it is clear that one can manipulate and transform the quantum state via any desired transformation, unitary or otherwise. It is also clear that such a 'brute force' approach to implementing gate operations is not very efficient. In the following section we describe a different and more effective strategy for addressing and manipulating individual qubits or pairs of qubits.

## 3. Projection operators

A basic step in quantum computation is the application of unitary or measurement gates to one or more specific qubits. Our approach to doing this will be to use projection operators to divide the state into projections onto the relevant subspaces corresponding to the addressed qubit(s). In this section, we first review the general theory of projections and then devise a scheme for constructing them for states defined by QMT signals.

### 3.1. General theory and notation

In order to address qubit A, a general two-qubit state may be written as

$$|\psi\rangle = |0\rangle_A \otimes \left[\alpha_{00}|0\rangle_B + \alpha_{01}|1\rangle_B\right] + |1\rangle_A \otimes \left[\alpha_{10}|0\rangle_B + \alpha_{11}|1\rangle_B\right] \quad (24)$$

$$= |0\rangle_A \otimes \left|\psi_0^{(A)}\right\rangle + |1\rangle_A \otimes \left|\psi_1^{(A)}\right\rangle. \quad (25)$$

Alternately, to address qubit B we may write

$$|\psi\rangle = |0\rangle_B \otimes \left[\alpha_{00}|0\rangle_A + \alpha_{10}|1\rangle_A\right] + |1\rangle_B \otimes \left[\alpha_{01}|0\rangle_A + \alpha_{11}|1\rangle_A\right] \quad (26)$$

$$= |0\rangle_B \otimes \left|\psi_0^{(B)}\right\rangle + |1\rangle_B \otimes \left|\psi_1^{(B)}\right\rangle. \quad (27)$$

Note that, although the order of the tensor products has been reversed, we use the subscript on the ket to keep track of which qubit it refers to, so there is no ambiguity of notation. The (unnormalized) two-qubit state $|0\rangle_B \otimes |\psi_0^{(B)}\rangle$, say, is the projection of $|\psi\rangle$ onto the subspace for which qubit B takes the value 0. The one-qubit state $|\psi_0^{(B)}\rangle$ shall be referred to as the *partial projection* and will play an important role in our QMT implementation.

For a general $n$-qubit state of the form given by equation (1), we may address a single qubit $i \in \{0, \ldots, n-1\}$ with value $a \in \{0, 1\}$ by defining the projection operator

$$\Pi_a^{(i)} |\psi\rangle = |a\rangle_i \otimes \left|\psi_a^{(i)}\right\rangle = \sum_{x:x_i=a} \alpha_x \left|x_{n-1}\cdots x_{i+1}\, a\, x_{i-1}\cdots x_0\right\rangle. \quad (28)$$

So, the partial projection operator $\pi_a^{(n,i)}: \otimes_{j=0}^{n-1} \mathcal{H}_j \to \otimes_{j=i+1}^{n-1} \mathcal{H}_j \otimes \overset{i-1}{\underset{j=0}{\otimes}} \mathcal{H}_j$ is defined to be $\pi_a^{(n,i)} |\psi\rangle = |\psi_a^{(i)}\rangle$. In the special case that $n = 1$, we have $\Pi_a^{(0)} |\psi\rangle = \alpha_a |a\rangle$ and $\pi_a^{(1,0)} |\psi\rangle = \alpha_a$. Note that $\Pi_0^{(i)} + \Pi_1^{(i)} = 1$ is the identity operator.

A two-qubit projection may be defined similarly. For $n \geqslant 2$, $i, j \in \{0, \ldots, n-1\}$, $a, b \in \{0, 1\}$, and $i \neq j$, we define the projection operator onto qubits $i$ and $j$ with values $a$ and $b$, respectively, as $\Pi_{ab}^{(ij)} := \Pi_b^{(j)} \Pi_a^{(i)} = \Pi_a^{(i)} \Pi_b^{(j)} = \Pi_{ba}^{(ji)}$, so

$$\Pi_{ab}^{(ij)} |\psi\rangle = |b\rangle_j \otimes |a\rangle_i \otimes \pi_b^{(n-1,j)} \pi_a^{(n,i)} |\psi\rangle = |a\rangle_i \otimes |b\rangle_j \otimes \left|\psi_{ab}^{(ij)}\right\rangle. \quad (29)$$

In the special case that $n = 2$, we have $|\psi_{ab}^{(10)}\rangle = \alpha_{ab} = |\psi_{ba}^{(01)}\rangle$. In the degenerate case that $i = j$, we define $\Pi_{ab}^{(ii)} = \Pi_a^{(i)}$ if $a = b$; otherwise, it is undefined. Note that, for $i \neq j$, $\Pi_{00}^{(ij)} + \Pi_{01}^{(ij)} + \Pi_{10}^{(ij)} + \Pi_{11}^{(ij)} = 1$ is the identity operator. The generalization to projections onto an arbitrary number of qubits is straightforward.

### 3.2. QMT implementation

Given an $n$-qubit state $|\psi\rangle$, as represented by a QMT complex basebanded signal $\psi$, we seek to construct the $n$-qubit state given by some projection $\Pi_0^{(i)} |\psi\rangle = |0\rangle_i \otimes |\psi_0^{(i)}\rangle$, say. We can see that this corresponds to a complex basebanded signal of the form $e^{i\omega_i t} \psi_0^{(i)}(t)$, so our task is to determine the partial projection signal $\psi_0^{(i)}(t)$. Of course, one can always do this in a brute-force manner by decomposing $\psi(t)$ into its $2^n$ complex components and then reconstructing the projection from these pieces. Our true task is to find a better construction scheme, one that does not require complete knowledge of the quantum state but relies only on the fact that it is a state of $n$ qubits.

*3.2.1. Single-qubit addressing*

Consider the one-qubit case $\psi(t) = \alpha_0 e^{i\omega t} + \alpha_1 e^{-i\omega t}$. If $n = 1$ and $i = 0$, then $\psi_0^{(0)}(t) = \alpha_0$ is a constant (DC) signal (i.e., a zero-qubit state). This may be constructed in the usual way by multiplying $\psi(t)$ by $\phi_0^\omega(t)^* = e^{-i\omega t}$ and low-pass filtering. In fact, this is nothing more than the one-qubit inner product function described earlier. We may construct $\psi_1^{(0)}(t) = \alpha_1$ in a similar manner.

Now consider the two-qubit case with $\omega_1 = \omega_A$ and $\omega_0 = \omega_B$. To address qubit A, we can construct $\psi_0^{(A)}(t)$ by multiplying $\psi(t)$ by $\phi_0^{\omega_A}(t)^* = e^{-i\omega_A t}$ and low-pass filtering. To see this, note that

$$\phi_0^{\omega_A}(t)^* \psi(t) = \alpha_{00} e^{i\omega_B t} + \alpha_{01} e^{-i\omega_B t} + \alpha_{10} e^{-i(2\omega_A - \omega_B)t} + \alpha_{11} e^{-i(2\omega_A + \omega_B)t}$$
$$= \psi_0^{(A)}(t) + \left[ \alpha_{10} e^{-i(2\omega_A - \omega_B)t} + \alpha_{11} e^{-i(2\omega_A + \omega_B)t} \right]. \tag{30}$$

The first term has frequencies $\pm \omega_B$, as desired. The remaining terms have frequencies $-(2\omega_A \pm \omega_B)$. Since $\omega_A \geqslant 2\omega_B > 0$, we see that $2\omega_A \pm \omega_B \geqslant 4\omega_B - \omega_B = 3\omega_B$. Thus, by using a low-pass filter with a passband of $|\omega| < 3\omega_B$, we may eliminate these remaining terms to obtain $\psi_0^{(A)}(t)$. A similar approach may be used to construct $\psi_1^{(A)}(t) = \alpha_{10} e^{i\omega_B t} + \alpha_{11} e^{-i\omega_B t}$. Given $\psi_0^{(A)}(t)$ and $\psi_1^{(A)}(t)$, the corresponding projections are easily constructed by multiplying the former by $e^{i\omega_A t}$ and the latter by $e^{-i\omega_A t}$.

To address qubit B, things are slightly more complicated. Multiplying $\psi(t)$ by $\phi_0^{\omega_B}(t)^*$ yields

$$\phi_0^{\omega_B}(t)^* \psi(t) = \alpha_{00} e^{i\omega_A t} + \alpha_{10} e^{-i\omega_A t} + \alpha_{01} e^{i(\omega_A - 2\omega_B)t} + \alpha_{11} e^{-i(\omega_A + 2\omega_B)t} \tag{31}$$

$$= \psi_0^{(0)}(t) + \left[ \alpha_{01} e^{i(\omega_A - 2\omega_B)t} + \alpha_{11} e^{-i(\omega_A + 2\omega_B)t} \right]. \tag{32}$$

The first term has frequencies $\pm \omega_A$, as desired. The remaining terms have frequencies $\omega_A - 2\omega_B \geqslant 0$ and $-(\omega_A + 2\omega_B) < 0$. Since $\omega_A \geqslant 2\omega_B > 0$, we see that $0 \leqslant \omega_A - 2\omega_B < \omega_A < \omega_A + 2\omega_B$. Thus, by using a bandpass filter with a passband of $|\omega - \omega_A| < 2\omega_B$, we may eliminate these terms to obtain $|\psi_0^{(B)}\rangle$. A similar approach may be used to construct $\psi_1^{(B)}(t) = \alpha_{01} e^{i\omega_A t} + \alpha_{11} e^{-i\omega_A t}$. Again, multiplication of $\psi_0^{(B)}(t)$ by $e^{i\omega_B t}$ and $\psi_1^{(B)}(t)$ by $e^{-i\omega_B t}$ yield the corresponding projections.

Now suppose $n > 2$. For this case, $i = n - 1$ is easy, $i = 0$ is as before, and $0 < i < n - 1$ is complicated. We begin, as before, by multiplying $\psi(t)$ by either $\phi_0^{\omega_i}(t)^* = e^{-i\omega_i t}$, to construct $\psi_0^{(i)}(t)$, or $\phi_1^{\omega_i}(t)^* = e^{i\omega_i t}$, to construct $\psi_1^{(i)}(t)$. For the former case, we have

$$\phi_0^{\omega_i}(t)^* \psi(t) = \sum_{x=0}^{2^n - 1} \alpha_x e^{i(\Omega_x - \omega_i)t} = \sum_{x: x_i = 0} \alpha_x e^{i(\Omega_x - \omega_i)t} + \sum_{x: x_i = 1} \alpha_x e^{i(\Omega_x - \omega_i)t}, \tag{33}$$

where $x_i = \mathrm{mod}[\mathrm{floor}(x/2^i), 2]$ is the value of bit $i$ in the binary expansion $x = x_0 2^0 + \cdots + x_{n-1} 2^{n-1}$ and

$$\Omega_x := (-1)^{x_{n-1}} \omega_{n-1} + \cdots + (-)^{x_0} \omega_0. \tag{34}$$

Note that the individual frequencies may be written in the form

$$\Omega_x - \omega_i = \begin{cases} \left[\Omega_x - (-1)^{x_i} \omega_i\right] & \text{if } x_i = 0 \\ \left[\Omega_x - (-1)^{x_i} \omega_i\right] - 2\omega_i & \text{if } x_i = 1. \end{cases} \tag{35}$$

From this, we may deduce that

$$\psi_0^{(i)}(t) = \sum_{x: x_i = 0} \alpha_x e^{i(\Omega_x - \omega_i)t}. \tag{36}$$

To actually construct this signal, we must bandpass filter the $2^{n-1}$ frequencies $\Omega_x - \omega_i$ corresponding to $x_i = 0$ or, equivalently, bandstop the $2^{n-1}$ frequencies $\Omega_x - \omega_i$ corresponding to $x_i = 1$. In the octave spacing scheme, the frequencies are $\Omega_x - \omega_i = (2^n - 1 - 2x - 2^i)\Delta\omega$ for $x = 0, \ldots, 2^n - 1$, of which only those for which $x_i = 0$ are bandpassed. Once this is done, we can multiply the bandpassed signal, $\psi_0^{(i)}(t)$, by $e^{i\omega_i t}$ to obtain the projection signal. Note that the state $\psi$ need not be known in order to construct this filter—only the total number of qubits, $n$, and the particular qubit to be addressed, $i$, need be specified.

A similar procedure may be used to obtain $\psi_1^{(i)}(t)$ and its corresponding projection. In this case, we multiply $\psi(t)$ by $\phi_1^{(i)}(t)^* = e^{i\omega_i t}$ to obtain

Table 1. Table of bandpass frequencies using the octave spacing scheme $\Omega_x \pm \omega_i = (2^n \pm 2^i - 1 - 2x)\Delta\omega$ for $x = 0, \ldots, 2^n - 1$. The '−' is for $x_i = 0$ and $\psi_0^{(i)}(t)$, while the '+' is for $x_i = 1$ and $\psi_1^{(i)}(t)$, although the frequencies are the same for both cases.

| $n$ | $i$ | $(\Omega_x \pm \omega_i)/\Delta\omega$ |
|---|---|---|
| 1 | 0 | 0 |
| 2 | 0 | −2  2 |
| 2 | 1 | −1  1 |
| 3 | 0 | −6 −2  2  6 |
| 3 | 1 | −5 −3  3  5 |
| 3 | 2 | −3 −1  1  3 |

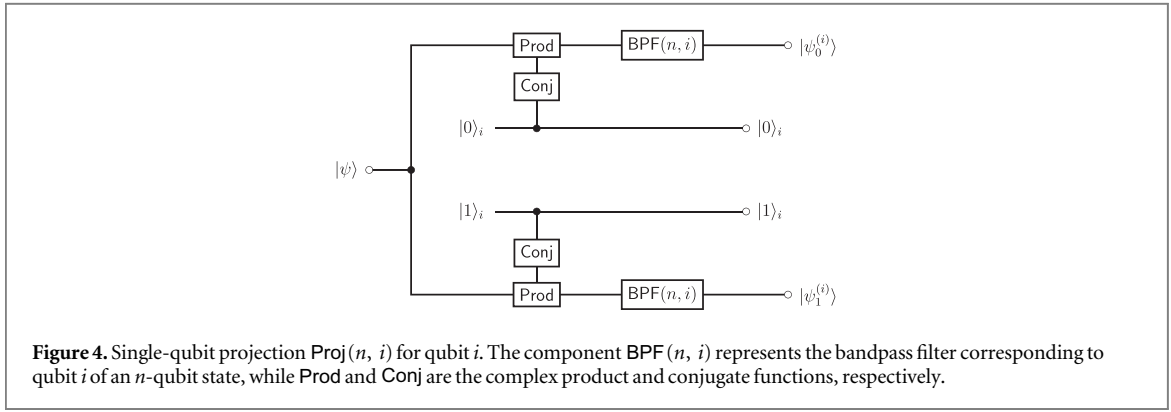

**Figure 4.** Single-qubit projection $\mathsf{Proj}(n, i)$ for qubit $i$. The component $\mathsf{BPF}(n, i)$ represents the bandpass filter corresponding to qubit $i$ of an $n$-qubit state, while $\mathsf{Prod}$ and $\mathsf{Conj}$ are the complex product and conjugate functions, respectively.

$$\psi_1^{(i)}(t) = \sum_{x:x_i=1} \alpha_x e^{i(\Omega_x + \omega_i)t}. \tag{37}$$

To construct this signal, we must therefore bandpass frequencies $\Omega_x + \omega_i$ such that $x_i = 1$. As it turns out, these are the same frequencies as $\Omega_x - \omega_i$ for $x_i = 0$. Table 1 gives the bandpass frequencies (in units of $\Delta\omega$) for $n = 1, 2, 3$ using an octave spacing scheme. A circuit schematic of the projection process is given in figure 4.

*3.2.2. Dual-qubit and multi-qubit addressing*
To address two distinct qubits, we need to construct a set of four partial projection signals of the form $\psi_{ab}^{(ij)}$ for $a, b \in \{0, 1\}$ from the single QMT signal $\psi$. This may be done by first projecting onto the $a$ subspace of qubit $i$ to obtain $\psi_a^{(i)}$, as described above, and then applying this same procedure to project onto the $b$ subspace of qubit $j$. A circuit schematic of the two-qubit projection process is given in figure 5.

The generalization to multi-qubit addressing is straightforward. In particular, if we project onto all $n$ subspaces, then

$$\pi_{x_0}^{(1,0)} \pi_{x_1}^{(2,1)} \cdots \pi_{x_{n-2}}^{(n-1,n-2)} \pi_{x_{n-1}}^{(n,n-1)} |\psi\rangle = \langle x_{n-1} \ldots x_0 | \psi \rangle = \alpha_x. \tag{38}$$

Note that each partial project involves only the use of a low-pass filter. This will be useful when we later consider measurement gates.

**3.3. Convolution as a form of bandpass filter**
The above projection scheme requires the use of comb-like bandpass filters to construct the partial projection signals. An alternative, yet equivalent, scheme would be to convolve the state $\psi$ with a *template* signal having the desired passband frequency components. The convolution of two such signals corresponds, in the frequency domain, to the product of their Fourier transforms Thus, in such a scheme, the template signal would serve as a mask to eliminate undesired frequency components. In this way, simple analog devices such as Surface Acoustic

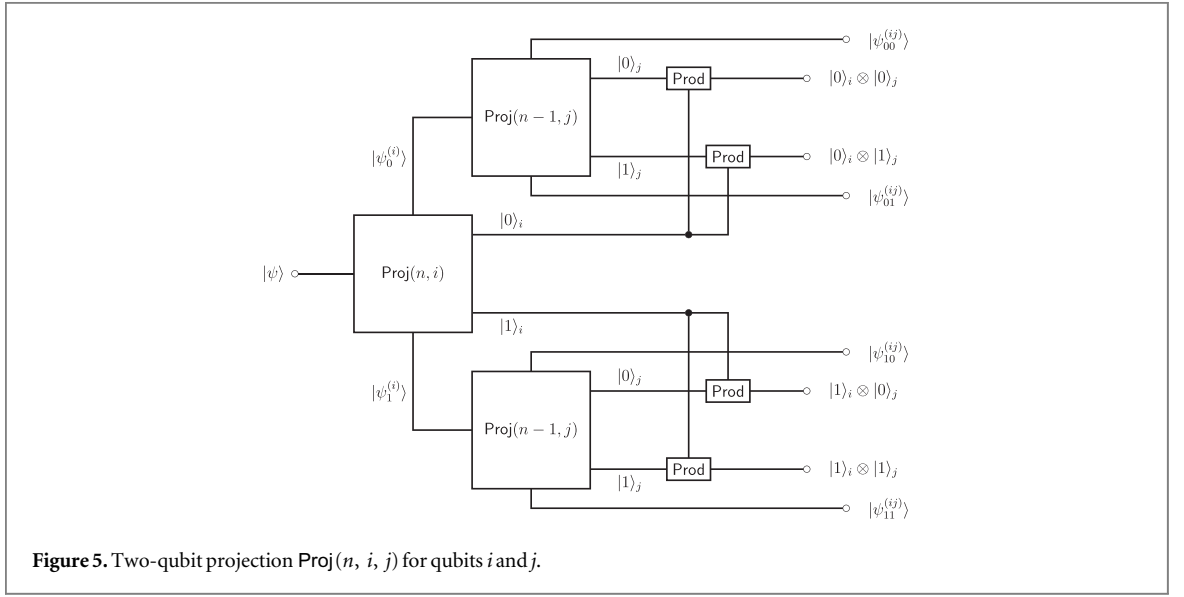

**Figure 5.** Two-qubit projection $\text{Proj}(n, i, j)$ for qubits $i$ and $j$.

Wave (SAW) elastic convolvers or optical charge-coupled device (CCD) convolvers may be used to construct programmable bandpass filters for each qubit [18, 19].

Suppose we have an $n$-qubit state represented by $\psi$ and wish to construct the partial projection state $\psi_0^{(i)}$, say, for qubit $i$. Using phase shifters and multipliers, we may easily construct a template signal of the form

$$\phi^{(i)}(t) := 2^{n-1} \cos(\omega_{n-1} t) \cdots \cos(\omega_{i+1} t) \cdot \cos(\omega_{i-1} t) \cdots \cos(\omega_0 t). \tag{39}$$

Note that the Fourier transform of $\phi^{(i)}$ is

$$\Phi^{(i)}(\omega) = \int_{-\infty}^{\infty} \phi^{(i)}(t) e^{-i\omega t} dt = 2\pi \sum_{y: y_i = 0} \delta(\omega - \Omega_y + \omega_i). \tag{40}$$

Now, consider the Fourier transform of $\hat{\psi}_0^{(i)}(t) := e^{-i\omega_i t} \psi(t) \frac{1}{T} 1_{[-T/2, T/2]}(t)$, given by

$$\hat{\Psi}_0^{(i)}(\omega) = \sum_{x=0}^{2^n - 1} \alpha_x \, \text{sinc}\left[(\omega - \Omega_x + \omega_i) T/2\right], \tag{41}$$

where sinc is the unnormalized sinc function. The convolution of $\phi^{(i)}$ and $\hat{\psi}_0^{(i)}$ is therefore

$$\left(\hat{\psi}_0^{(i)} \star \phi^{(i)}\right)(t) = \sum_{x=0}^{2^n - 1} \sum_{y: y_i = 0} \alpha_x \, \text{sinc}\left[(\Omega_y - \Omega_x) T/2\right] e^{i(\Omega_y - \omega_i) t}. \tag{42}$$

For $\Omega_y = \Omega_x$ the argument of the sinc function is zero and, hence, takes on a value of 1. For the remaining terms the sinc function become vanishingly small as $T$ becomes large. In the special case of the octave spacing scheme and an integer number of periods [i.e., $T \in (2\pi/\omega_0)\mathbb{N}$], the argument of the sinc function becomes $(\Omega_y - \Omega_x) T/2 \in 2\pi(x - y)\mathbb{N}$. So, all terms such that $x \neq y$ drop out, leaving only those such that $x_i = 0$. The result is precisely the desired partial projection signal:

$$\left(\hat{\psi}_0^{(i)} \star \phi^{(i)}\right)(t) = \sum_{x: x_i = 0} \alpha_x e^{i(\Omega_x - \omega_i) t} = \psi_0^{(i)}(t). \tag{43}$$

A similar procedure may be followed to obtain other qubit projection states.

## 4. Gate operations

Using the results of the previous section, we may operate on the state $|\psi\rangle$, represented by the complex basebanded signal $\psi$, by decomposing it into projection signals and operating on each component.

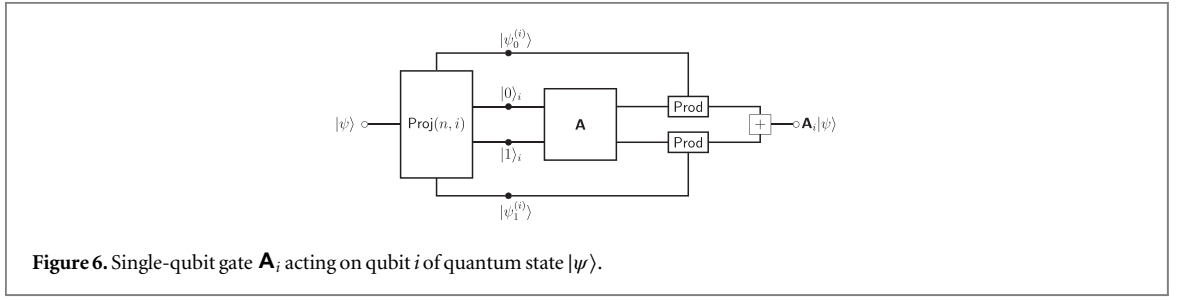

**Figure 6.** Single-qubit gate $\mathbf{A}_i$ acting on qubit $i$ of quantum state $|\psi\rangle$.

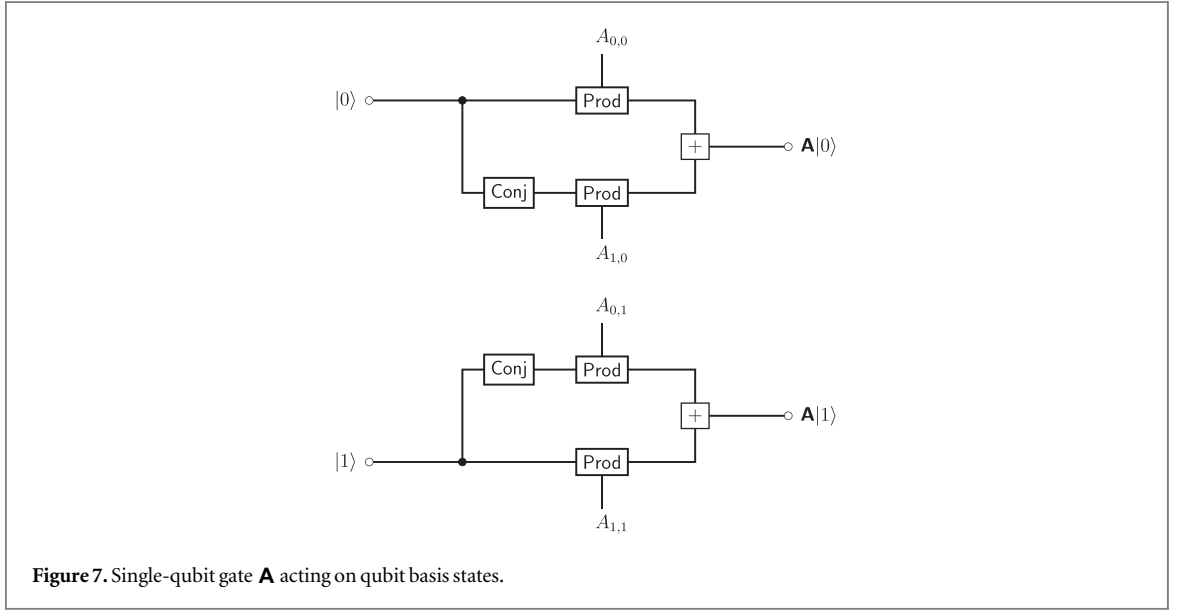

**Figure 7.** Single-qubit gate $\mathbf{A}$ acting on qubit basis states.

A linear operation on a single qubit may be represented by a complex $2 \times 2$ matrix $\mathbf{A}$, where

$$\mathbf{A} = \begin{pmatrix} A_{0,0} & A_{0,1} \\ A_{1,0} & A_{1,1} \end{pmatrix}. \tag{44}$$

If $\mathbf{A}$ acts on qubit $i$ of state $|\psi\rangle$, then we write

$$\mathbf{A}_i \,|\psi\rangle = \left[ A_{0,0} \,|0\rangle_i + A_{1,0} \,|1\rangle_i \right] \otimes \left| \psi_0^{(i)} \right\rangle + \left[ A_{0,1} \,|0\rangle_i + A_{1,1} \,|1\rangle_i \right] \otimes \left| \psi_1^{(i)} \right\rangle. \tag{45}$$

A schematic illustration of the process is shown in figure 6 using a linear gate $\mathbf{A}$. The construction of the gate itself is illustrated in figure 7.

For example, the action of a NOT gate $\mathbf{X}$, where

$$\mathbf{X} = \begin{pmatrix} 0 & 1 \\ 1 & 0 \end{pmatrix}, \tag{46}$$

on qubit 0 of a two-qubit state would be

$$\begin{aligned} \mathbf{X}_0 \,|\psi\rangle &= |1\rangle_0 \otimes \left| \psi_0^{(0)} \right\rangle + |0\rangle_0 \otimes \left| \psi_1^{(0)} \right\rangle \\ &= |1\rangle_0 \otimes \left[ \alpha_{00} \,|0\rangle_1 + \alpha_{10} \,|1\rangle_1 \right] + |0\rangle_0 \otimes \left[ \alpha_{01} \,|0\rangle_1 + \alpha_{11} \,|1\rangle_1 \right] \\ &= \alpha_{00} \,|01\rangle + \alpha_{01} \,|00\rangle + \alpha_{10} \,|11\rangle + \alpha_{11} \,|10\rangle. \end{aligned} \tag{47}$$

Similarly, if $\mathbf{B}$ is a two-qubit operator of the form

$$\mathbf{B} = \begin{bmatrix} B_{00,00} & B_{00,01} & B_{00,10} & B_{00,11} \\ B_{01,00} & B_{01,01} & B_{01,10} & B_{01,11} \\ B_{10,00} & B_{10,01} & B_{10,10} & B_{10,11} \\ B_{11,00} & B_{11,01} & B_{11,10} & B_{11,11} \end{bmatrix} \tag{48}$$

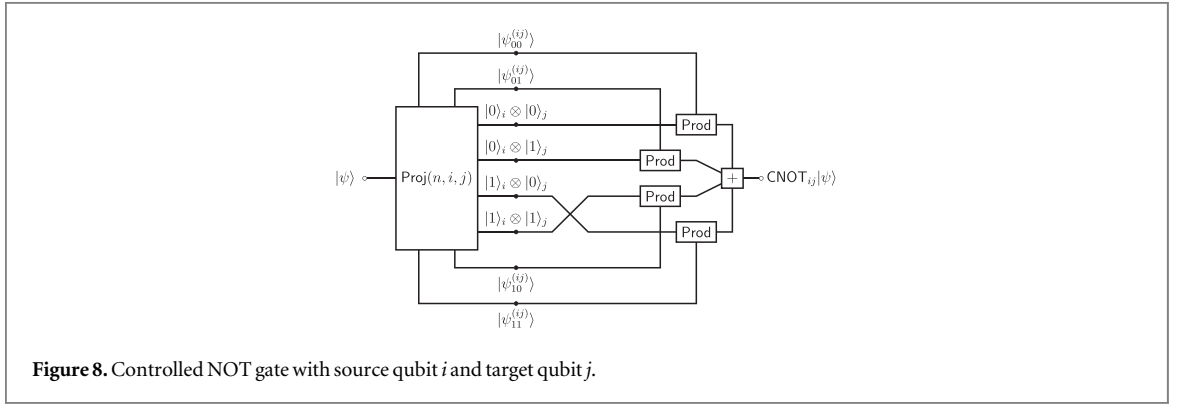

**Figure 8.** Controlled NOT gate with source qubit *i* and target qubit *j*.

acting on qubits *i* and *j*, then we may write

$$\mathbf{B}_{ij} |\psi\rangle = \sum_{a,b=0}^{1} \left[ \sum_{a',b'=0}^{1} B_{a'b',ab} |a'\rangle_i \otimes |b'\rangle_j \right] \otimes \left|\psi_{ab}^{(ij)}\right\rangle. \quad (49)$$

For example, the action of a controlled NOT (CNOT) gate **C**, where

$$\mathbf{C} = \begin{bmatrix} \mathbf{I} & \mathbf{0} \\ \mathbf{0} & \mathbf{X} \end{bmatrix}, \quad (50)$$

acting on qubits *i* and *j*, with *i* the controlling or source qubit and *j* the controlled or target qubit, would be

$$\begin{aligned}
\mathbf{C}_{ij} |\psi\rangle &= \mathbf{C}_{ij} \left[ |0\rangle_i \otimes \pi_0^{(n,i)} |\psi\rangle + |1\rangle_i \otimes \pi_1^{(n,i)} |\psi\rangle \right] \\
&= |0\rangle_i \otimes \left|\psi_0^{(i)}\right\rangle + \mathbf{C}_{ij} |1\rangle_i \otimes \left[ |0\rangle_j \otimes \left|\psi_{10}^{(ij)}\right\rangle + |1\rangle_j \otimes \left|\psi_{11}^{(ij)}\right\rangle \right] \\
&= |0\rangle_i \otimes \left|\psi_0^{(i)}\right\rangle + |1\rangle_i \otimes \left[ |1\rangle_j \otimes \left|\psi_{10}^{(ij)}\right\rangle + |0\rangle_j \otimes \left|\psi_{11}^{(ij)}\right\rangle \right].
\end{aligned} \quad (51)$$

Figure 8 illustrates one possible schematic for implementing this gate.

For the application of any linear operator, the same general procedure is followed: first, we construct the partial projection signals corresponding to the qubit(s) to which we wish to apply the operator. Next, the remaining basis states are then transformed according to the operator matrix elements. Finally, these signals are multiplied to the partial projection signals to obtain the transformed state.

The above procedures describe how to apply general *linear* transformations to the QMT state. This of course includes, but is not limited to, *unitary* transformations. For this reason, and the simple fact that one may copy a classical signal, the No Cloning theorem does not apply. Of course, the theorem still applies if we restrict ourselves to using only unitary gates, but this restriction is not necessary.

Going one step further, there is no reason why one cannot apply *nonlinear* transformations to the QMT state, and such a possibility has very interesting implications. According to a well-known result from Abrahams and Lloyd [20], the ability to perform nonlinear transformations in a quantum computer allows one to solve oracle-based **#P** and **NP**-complete problems in polynomial time. This result has thus far been of mere theoretical interest, as quantum mechanics appears to be stubbornly linear, but our approach suggests that it may yet be of some practical utility.

## 5. Measurement gates

Our final task is to extract information from the signal $\psi$ representing the quantum state $|\psi\rangle$. One approach would be to perform a full analysis of all complex components $\alpha_x$. This is an order $2^n$ procedure in the number of required operations, which we shall call the *Brute Force* approach. For true quantum systems, of course, this cannot be done. Instead, information must be extracted by measuring individual qubits, thereby obtaining a binary sequence and a projection of the state according to the particular sequence of outcomes. Performing sequential measurements of this sort is an order $n$ procedure.

Suppose we wish to measure qubit *i*. We begin by constructing the projections $\Pi_0^{(i)} |\psi\rangle$ and $\Pi_1^{(i)} |\psi\rangle$ from the partial projections $|\psi_0^{(i)}\rangle$ and $|\psi_1^{(i)}\rangle$, respectively. Let

$$q_0^{(i)} := \|\Pi_0^{(i)} |\psi\rangle\|^2 = \langle\psi| \Pi_0^{(i)} |\psi\rangle, \quad q_1^{(i)} := \|\Pi_1^{(i)} |\psi\rangle\|^2 = \langle\psi| \Pi_1^{(i)} |\psi\rangle \tag{52}$$

denote the magnitudes of these projections. Then, according to the generalized Born rule, the probability of outcome $a \in \{0, 1\}$ is

$$p_a^{(i)} := \frac{\langle\psi| \Pi_a^{(i)} |\psi\rangle}{\langle\psi|\psi\rangle} = \frac{q_a^{(i)}}{q_0^{(i)} + q_1^{(i)}}, \tag{53}$$

where we recall that $|\psi\rangle$ is not assumed to be normalized.

The quantum state after measurement, in accordance with the projection postulate of wavefunction collapse, will be one of the two projections (either $\Pi_0^{(i)} |\psi\rangle$ or $\Pi_1^{(i)} |\psi\rangle$), depending upon which outcome is obtained. (Again, we do not insist upon normalized states.) Equivalently, we may take the collapsed state to be one of the partial projections (either $|\psi_0^{(i)}\rangle$ or $|\psi_1^{(i)}\rangle$), thereby collapsing to a state of $n - 1$ qubits.

To measure a second qubit $j \neq i$, the same procedure is followed. From the collapsed state $|\psi'\rangle = \Pi_a^{(i)} |\psi\rangle$ we compute the conditional weights

$$q_{b|a}^{(j|i)} := \langle\psi'| \Pi_b^{(j)} |\psi'\rangle = \langle\psi| \Pi_a^{(i)} \Pi_b^{(j)} \Pi_a^{(i)} |\psi\rangle = \langle\psi| \Pi_a^{(i)} \Pi_b^{(j)} |\psi\rangle \tag{54}$$

for $b \in \{0, 1\}$. The conditional probability of obtaining outcome $b$ on qubit $j$, given outcome $a$ on qubit $i$, is therefore

$$p_{b|a}^{(j|i)} := \frac{\langle\psi| \Pi_a^{(i)} \Pi_b^{(j)} |\psi\rangle}{\langle\psi| \Pi_a^{(i)} |\psi\rangle} = \frac{q_{b|a}^{(j|i)}}{q_a^{(i)}}. \tag{55}$$

The joint probability of both outcomes is then $p_{ab}^{(ij)} := p_a^{(i)} p_{b|a}^{(j|i)} = \langle\psi| \Pi_a^{(i)} \Pi_b^{(j)} |\psi\rangle / \langle\psi|\psi\rangle$.

In the QMT representation we have developed thus far, there is no sense of a random outcome; whereas, a quantum computer is generally conceived as being probabilistic. There are several approaches one might take, then, to representing a measurement gate. One approach, the Brute Force approach described earlier, would be to extract explicity all $2^n$ complex components of the state, a process that, as we have noted, is not very efficient. We therefore consider the following three alternative approaches.

*Binary search*: in many quantum computer algorithms, the final state has either a single non-zero component or one dominant component, corresponding to the correct answer. In this case, an order-$n$ scaling procedure may be used to compute sequentially first $q_0^{(0)}$ and $q_1^{(0)}$, then either $q_{0|0}^{(0|0)}$ and $q_{1|0}^{(1|0)}$ (if $q_0^{(0)} > q_1^{(0)}$) or $q_{0|1}^{(0|1)}$ and $q_{1|1}^{(1|1)}$ (otherwise), etc, selecting the larger of the two at each stage and thereby perform a binary search to identify the non-zero (or dominant) component.

*Simulation*: with a ready supply of random numbers serving the role of hidden variables, one could replicate the generalized Born rule and thereby replicate quantum statistics rather trivially. Suppose $u_0, u_1, \ldots, u_{n-1}$ is a realization of $n$ independent and identically distributed uniform random variables drawn from the unit interval [0, 1]. We may specify that the outcome of measuring qubit 0, denoted $x_0$, is 1 if $u_0 > p_0^{(0)}$ and is 0 otherwise. The signal $\psi$ is then replaced by that corresponding to the collapsed projection $\Pi_{x_0}^{(0)} |\psi\rangle$. Continuing in this manner, a measurement of qubit $i$, given the previous outcomes $x_0, \ldots, x_{i-1}$, yields $x_i = 1$ if $u_i > p_{0|x_{i-1},\ldots,x_0}^{(i|i-1,\ldots,0)} = \langle\psi| \Pi_0^{(i)} \Pi_{x_{i-1}}^{(i-1)} \cdots \Pi_{x_0}^{(0)} |\psi\rangle / \langle\psi| \Pi_{x_{i-1}}^{(i-1)} \cdots \Pi_{x_0}^{(0)} |\psi\rangle$ and $x_i = 0$ otherwise for $i = 0, \ldots, n - 1$. This approach may be useful for quantum simulation. For the case in which there is one dominant component, this approach reduces to a binary search.

*Threshold detection*: a fourth, and much more ambitious approach, would be to use a wholly deterministic formulation of the measurement procedure that nevertheless reproduces, or at least approximates, the generalized Born rule. One example would be to use a signal-plus-noise model with amplitude threshold detection. In this scheme, each $\alpha_x$ is replaced by $a_x = s\alpha_x + \nu_x$, where $s \geq 0$ is a scale factor and each $\nu_x$ is a complex random variable representing a hidden variable state. Collectively, $\nu_0, \ldots, \nu_{2^n-1}$ follow a certain joint probability distribution. For example, they may be independent and identically distributed complex Gaussian random variables with zero mean. The quantum state given by equation (1) is now represented by a *random signal* of the form

$$\psi(t) = \sum_{x=0}^{2^n-1} a_x \phi_x(t). \tag{56}$$

The weights $q_0^{(i)} = \langle\psi| \Pi_0^{(i)} |\psi\rangle$ and $q_1^{(i)} = \langle\psi| \Pi_1^{(i)} |\psi\rangle$, now random variables by virtue of the noise term, are computed in the same manner but now are compared against a threshold $\gamma^2 > 0$ for a particular realization of $\psi$. If exactly one of the two components exceeds the threshold, then we say that a single detection has been made (i.e., a valid measurement has been performed) and the outcome is either 0 or 1, depending upon whether $q_0^{(i)} > \gamma^2$ or $q_1^{(i)} > \gamma^2$, respectively. Multi-qubit measurements use wavefunction collapse to the detected subspace projection, as described above. Surprisingly, for many cases of interest to quantum computing, it can be shown that such an approach yields results that are comparable, if not identical, to those predicted by the Born rule [16].

## 6. Quantum teleportation using QMT

To illustrate the application of the QMT representation to quantum computing, we consider here an example of quantum teleportation, a quantum effect first described in 1993 by Bennett *et al* [21] and experimentally realized by Bouwmeester *et al* in 1997 [22]. The protocol concerns a three-qubit system such that an initial state of the form

$$|\psi\rangle = |\phi\rangle_A \otimes \frac{|0\rangle_S \otimes |0\rangle_B + |1\rangle_S \otimes |1\rangle_B}{\sqrt{2}}, \tag{57}$$

where $|\phi\rangle_A = \alpha |0\rangle_A + \beta |1\rangle_A$ is an arbitrary one-qubit state, is transformed into a final state of the form

$$|\psi'\rangle = |x\rangle_A \otimes |y\rangle_S \otimes |\phi\rangle_B, \tag{58}$$

where $x, y \in \{0, 1\}$ and $|\phi\rangle_B = \alpha |0\rangle_B + \beta |1\rangle_B$.

As is common in such discussions, we shall describe the process in terms of a game between two familiar characters, Alice and Bob. The rules of the game are that Alice is able to manipulate her qubit (A) and the shared qubit (S) but not Bob's qubit (B). Bob, similarly, can manipulate B and S but not A. One further restriction, implicit in true quantum systems but imposed artificially for our QMT representation, is that Alice cannot simply analyze and copy the state $|\phi\rangle_A$ and then send it to Bob (as that would be cheating). The only valid actions are the application of unitary gates to the qubits to which they have access and classical communication.

For our QMT representation, we begin the story at a slightly earlier time, with Alice in possession only of her qubit, a complex basebanded signal of the form

$$\psi_A(t) = \alpha \phi_0^{\omega_A}(t) + \beta \phi_1^{\omega_A}(t), \tag{59}$$

and Bob in possession of qubits S and B, which take the form of separate signals $|0\rangle_S = \phi_0^{\omega_S}$ and $|0\rangle_B = \phi_0^{\omega_B}$. (Note that the two qubits are differentiated by their distinct frequencies, $\omega_S$ and $\omega_B$.) He multiplies these two signals, obtaining the separable state

$$\psi_B(t) = \phi_0^{\omega_S}(t) \phi_0^{\omega_B}(t), \tag{60}$$

and applies first a Hadamard gate $\mathbf{H}_S$ and then a CNOT gate $\mathbf{C}_{S,B}$, using S as the control and B as the target, to obtain an entangled Bell state given by

$$\psi'_B(t) = \frac{1}{\sqrt{2}} \Big[ \phi_0^{\omega_S}(t) \phi_0^{\omega_B}(t) + \phi_1^{\omega_S}(t) \phi_1^{\omega_B}(t) \Big]. \tag{61}$$

Finally, Bob sends this signal to Alice, who then multiplies it to her own qubit to obtain

$$\psi = \Big[ \alpha \phi_0^{\omega_A} + \beta \phi_1^{\omega_A} \Big] \cdot \frac{\phi_0^{\omega_S} \cdot \phi_0^{\omega_B} + \phi_1^{\omega_S} \cdot \phi_1^{\omega_B}}{\sqrt{2}}. \tag{62}$$

Note that Bob cannot send just the shared qubit, S. Because it is entangled with qubit B, he can only send the combined, two-qubit signal. This would appear to be the key difference between this classical representation and a true quantum system.

Following the standard teleportation protocol, Alice now applies first a CNOT gate $\mathbf{C}_{A,S}$ and then a Hadamard gate $\mathbf{H}_S$ on qubit S. As is well known, this results in the state

$$\frac{1}{2} \sum_{x,y=0}^{1} \phi_x^{\omega_A} \cdot \phi_y^{\omega_S} \cdot \mathbf{X}_B^x \mathbf{Z}_B^y \Big[ \alpha \phi_0^{\omega_B} + \beta \phi_1^{\omega_B} \Big], \tag{63}$$

where $\mathbf{Z} |0\rangle = |0\rangle$ and $\mathbf{Z} |1\rangle = -|1\rangle$. Note that, although Alice is in possession of Bob's qubit, she does not manipulate it. Nevertheless, her actions on qubits A and S result in the transfer of state information from her qubit to Bob's qubit.

The next step in the procedure calls for Alice to perform a measurement on qubits A and S. For the QMT state, this is done by first projecting the above state onto each of the four basis states for qubits A and S. This results in Alice now having four separate projection signals, each of the form

$$\phi_x^{\omega_A} \cdot \phi_y^{\omega_S} \cdot \mathbf{X}_B^x \mathbf{Z}_B^y \left[ \alpha \phi_0^{\omega_B} + \beta \phi_1^{\omega_B} \right]. \tag{64}$$

To complete the protocol, Alice, in lieu of a random measurement, may simply pick one of these four projections, send it to Bob, and communicate to him the chosen values of $x$ and $y$. Bob, in turn, uses this information to apply the inverse operation $\mathbf{Z}_B^y \mathbf{X}_B^x$ for the specified values of $x$ and $y$, thereby obtaining

$$\psi' = \phi_x^{\omega_A} \cdot \phi_y^{\omega_S} \cdot \left[ \alpha \phi_0^{\omega_B} + \beta \phi_1^{\omega_B} \right], \tag{65}$$

as desired.

Alternatively, Alice could separate the B qubit signal and send only this, and her two classical bits, to Bob. Bob, applying the inverse, would then be in possession of Alice's initial state, and Alice would be left holding two separable basis states, thereby reversing their initial roles in the story.

## 7. Quantum parallelism

One of the first demonstrations of the computational advantage of a quantum computer was given by David Deutsch [4]. The original problem concerns a simple Boolean function $f : \{0, 1\} \mapsto \{0, 1\}$ and determining whether it is such that $f(0) = f(1)$ or $f(0) \neq f(1)$. In the more general Deutsch–Jozsa problem, we wish to determine whether a given function $f : \{0, 1\}^n \mapsto \{0, 1\}$ is either *constant* (i.e., $f(x) = f(y)$ for all $x, y$) or *balanced* (i.e., $f(x) = 0$ for exactly half of the possible values of $x$), assuming it is one of the two [5]. On a classical (digital) computer, the only way to do this, with certainty, is to evaluate $f$ for up to $2^n/2 + 1$ inputs (in case the first half all give the same value). On a quantum computer, only a single application of $f$ is needed, due to quantum parallelism. How well can a classical *analog* computer do?

It has already been noted that the Deutsch–Jozsa algorithm may be implemented classically. For example, in 2011 Dyson demonstrated that one could implement the algorithm using only analog electronic circuits [23]. This particular approach relied upon the fact that the quantum state components remain real-valued in the Deutsch–Jozsa algorithm, so the method does not generalize to arbitrary quantum state operations. Furthermore, due to the explicit representation of matrix multiplication operations, this approach does not scale well with the number of qubits, although the execution time is constant.

Let us now follow the standard quantum computing algorithm, but using a QMT signal representation. We begin with a signal of the form

$$\psi(t) = e^{i\omega_n t} \cdots e^{i\omega_1 t} \, e^{-i\omega_0 t} \tag{66}$$

representing the $(n + 1)$-qubit quantum state $|\psi\rangle = |0 \cdots 0\rangle |1\rangle$. (The left $n$ qubits represent the input register, while the right qubit represents the output register.) In accordance with the quantum algorithm, we now apply $n$ Hadamard gates to the input register and one to the output register to obtain

$$\psi'(t) = i \, 2^{(n+1)/2} \cos(\omega_n t) \cdots \cos(\omega_1 t) \, \sin(\omega_0 t). \tag{67}$$

This provides a superposition of all $2^n$ possible inputs. More explicitly, one may write

$$\psi'(t) = i \sum_{x_n=0}^{1} \cdots \sum_{x_1=0}^{1} \frac{\phi_{x_n}^{\omega_n}(t) \cdots \phi_{x_1}^{\omega_1}(t)}{\sqrt{2^{n-1}}} \sin(\omega_0 t). \tag{68}$$

Next, we apply an $n$-qubit unitary gate $\mathbf{U}_f$ to the input register, and a Hadamard to the output register, such that the resulting signal is

$$\psi''(t) = \sum_{\mathbf{x} \in \{0,1\}^n} \frac{(-1)^{f(\mathbf{x})}}{\sqrt{2^n}} \phi_{x_n}^{\omega_n}(t) \cdots \phi_{x_1}^{\omega_1}(t) \, e^{-i\omega_0 t}, \tag{69}$$

where $\mathbf{x} = (x_1, \ldots, x_n)$. We know that such a $\mathbf{U}_f$ may be constructed from a polynomial number of one- and two-qubit gates, and the same is true for our signal using analog filters.

In particular, we may specify $f$ uniquely by a parameter $a \in \{0, \ldots, 2^{n+1} - 1\}$ whose (little endian) binary representation is written $a_n \cdots a_1 a_0$. In terms of $a$, then, $\mathbf{U}_f$ may be written explicitly as

$$\mathbf{U}_f = \mathbf{C}_{n0}^{a_n} \cdots \mathbf{C}_{10}^{a_1} \mathbf{X}_0^{a_0}, \tag{70}$$

where $\mathbf{C}_{i0}$ is a CNOT gate with control qubit $i$ and target qubit 0, $\mathbf{X}_0$ is a NOT gate applied to qubit 0, and a zero exponent corresponds to the identity. Note that $f$ is constant if and only if $a < 2$.

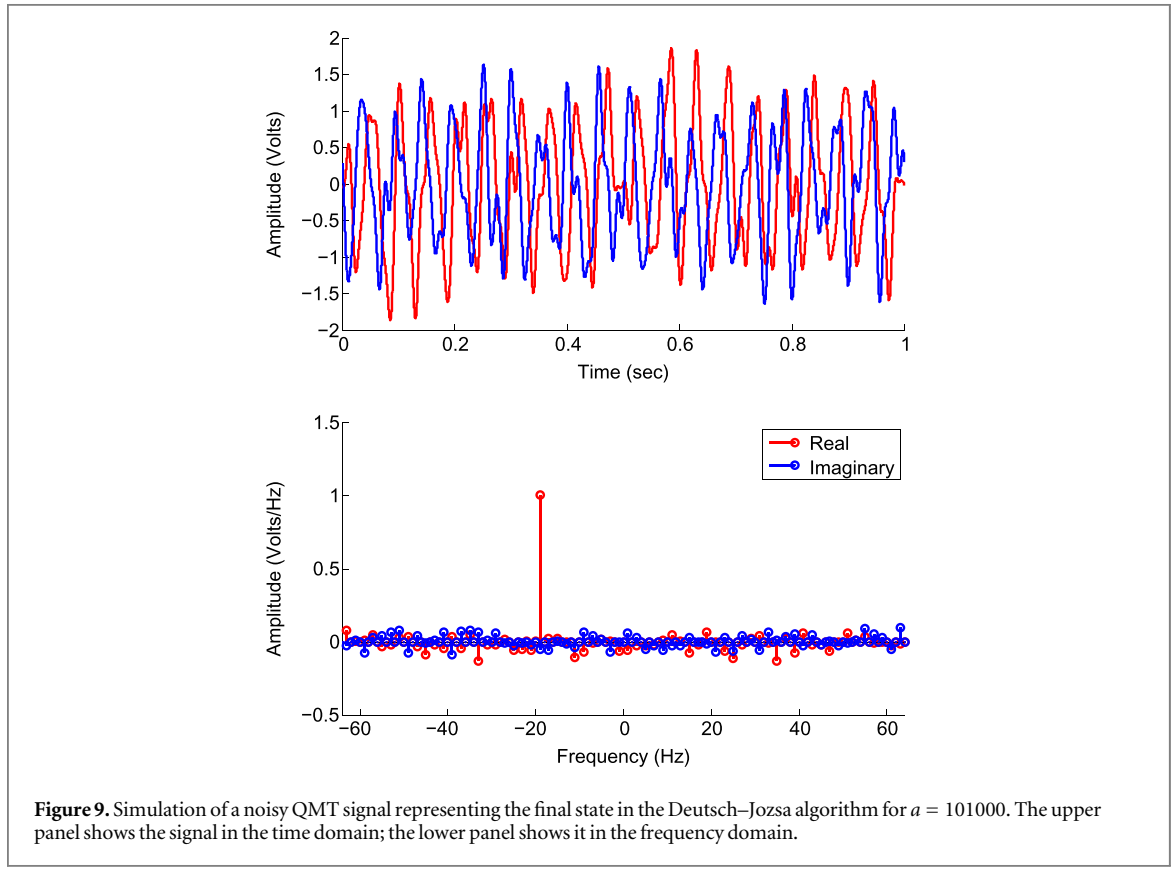

**Figure 9.** Simulation of a noisy QMT signal representing the final state in the Deutsch–Jozsa algorithm for $a = 101000$. The upper panel shows the signal in the time domain; the lower panel shows it in the frequency domain.

In the final step, we apply $n$ Hadamard gates to the input register to obtain

$$\psi'''(t) = \sum_{\mathbf{x},\mathbf{y} \in \{0,1\}^n} \frac{(-1)^{f(\mathbf{x})+\mathbf{x}\cdot\mathbf{y}}}{2^n} \phi_{y_n}^{\omega_n}(t) \cdots \phi_{y_1}^{\omega_1}(t) \, e^{-i\omega_0 t}, \quad (71)$$

where $\mathbf{x} \cdot \mathbf{y} = x_1 y_1 + \cdots + x_n y_n$ (modulo 2).

The resulting signal is such that there is always exactly one non-zero frequency component (i.e., the quantum state has exactly one non-zero amplitude). In particular, $f$ will be constant if and only if the component $|0\cdots 0\rangle|1\rangle$ of $|\psi'''\rangle$ has unit magnitude. To determine whether or not this is the case, one need only measure the input register. How does one do this efficiently with a signal?

Following our previous discussion on measurement gates, we can sequentially measure qubits $n$ through 1 by first multiplying $\psi'''(t)$ by $e^{-i\omega_n t}$, low-pass filtering the resulting signal, and then computing the magnitude of the resulting projection state. Continuing in this manner allows an efficient binary search over the $n$ input register qubits. If the outcome of *any* of these measurements is not 0, then we conclude that $f$ is balanced; otherwise, we conclude that it is constant.

Interestingly, one can go further and extract the actual value of the parameter $a$. It can be shown that the first $n$ binary digits of $a$ (i.e., $a_n, \ldots, a_1$) are given the measurement outcomes of qubits $n$ through 1, respectively. Now, by construction, the qubit 0 state is $|1\rangle$, independent of $a_0$; however, if we examine the *sign* of the non-zero frequency component (a thing not possible in a true quantum system), we may deduce that $a_0 = 0$ if it is positive and $a_0 = 1$ if it is negative. Specifically, using the deduced values of the first $n$ qubits, we may compute the inner product $\langle a_n \cdots a_1 1 | \psi''' \rangle$. If (the real part of) this inner product is positive, then $a_0 = 0$; otherwise, $a_0 = 1$. In this way, we may not only determine whether $f$ is constant or balanced but uniquely identify which of the $2^{n+1}$ possible functions was implemented using only a single application of the oracle!

This algorithm was implemented in Matlab using a digital simulation of the signals and gate operations described in this paper. Idealized filters and lossless components were assumed in the simulation. To study robustness, white Gaussian noise was added to the input signal. Figure 9 shows the results of one such run of the Deutsch–Jozsa algorithm for $n = 5$. The figure shows the final signal, prior to measurement, with noise added to the signal to achieve a $-10$ dB signal-to-noise ratio (SNR). Despite the low SNR value, the simulated measurements were easily able to correctly estimate the function parameter value ($a = 101000$, in this case).

## 8. Discussion

Having developed the basic theory behind the QMT representation, we now turn to some practical considerations and computational implications of this approach.

### 8.1. Practical implementation

In theory, the QMT representation may be used for an arbitrary number of qubits. From practical considerations, the number of qubits that can be represented will be limited by the bandwidth and spectral resolution that can actually be achieved. For electronic signals, frequencies in the range of 0.1 Hz to 100 GHz can be reasonably accommodated with standard components. Spectral resolution will be dictated by the signal duration $T$, scaling roughly as $1/T$. Thus, with $T \sim 10$ sec, this corresponds to a frequency resolution of about $\Delta f = 1/T \sim 0.1$ Hz. This suggests that a single signal of modest duration could accommodate as many as 40 qubits. This is comparable to a modern high performance computer, which has about 1 TB of RAM and can therefore store about $2^{42}$ double precision floating point numbers (i.e., about $2^{41}$ complex numbers). Thus, in terms of memory capacity alone this rivals current state-of-the-art digital technology. Being an analog system, however, processing speeds can be expected to be much faster (on the order of the signal duration), as the system architecture is intrinsically parallel.

Larger values of $T$ allow for smaller values of $\Delta f$ and, hence, larger values of $n$. For $T \sim 10$ hr one could have 50 qubits, while for $T \sim 1$ yr one could have 60 qubits. Even if $T$ were on the order of the age of the Universe (about 13.77 billion years), though, one could only represent about 95 qubits. Pushing on the other side of the spectrum, down to the Planck time scale (about $10^{-43}$ s), brings the number of qubits up to only about 176. This still falls well short of the *thousands* of qubits needed for a practical implementation of, say, Shor's algorithm for breaking RSA encryption. Nevertheless, even several tens of qubits may still be of significant practical benefit.

The signal representing the quantum state must be of an amplitude that is large enough to discriminate the various frequency components from background noise yet small enough to be practical. For electronic circuits operating in the aforementioned frequency range, background noise consists mainly of resistor thermal (or Johnson–Nyquist) noise. Thermal noise can be modeled as complex additive white Gaussian noise, characterized by the power spectral density

$$\sigma^2 = 4k_B TR, \quad (72)$$

where $k_B = 1.38 \times 10^{-23}$ J/K is Boltzmann's constant, $T$ is the ambient temperature (in K), and $R$ is the load resistance (in $\Omega$). For typical values of $T = 300$ K and $R = 10$ k$\Omega$, this corresponds to a power spectral density of about $\sigma^2 \approx (13\,\mu V)^2/$Hz. With a spectral resolution of about 0.1 Hz, this corresponds to an RMS value of about $\sigma\sqrt{\Delta f} = 4\,\mu$V per spectral component.

In the presence of noise, our complex, basebanded signal becomes

$$\psi(t) + w(t), \quad (73)$$

where $w$ is a complex, zero-mean white noise process such that $E[w(t)w^\star(t')] = \sigma^2 \delta(t-t')$. A quantum state with random components may be considered a mixed state. In particular, it can be shown that this random signal corresponds to a mixed state of the form

$$\rho \propto |\psi\rangle\langle\psi| + \sigma^2 \Delta f\, \mathbf{I}, \quad (74)$$

where $\mathbf{I}$ is the identity. The presence of additive white noise therefore has the effect of a depolarization channel $\mathcal{E}(\rho) = (1-p)\rho + (p/N)\mathbf{I}$ with a mixing ratio of

$$p = \left(1 + \frac{\|\psi\|^2}{N\sigma^2 \Delta f}\right)^{-1}. \quad (75)$$

The presence of thermal noise, then, corresponds to a quantum state fidelity of

$$F = \frac{\sqrt{\langle\psi|\rho|\psi\rangle}}{\|\psi\|} = \sqrt{\frac{\|\psi\|^2 + \sigma^2 \Delta f}{\|\psi\|^2 + N\sigma^2 \Delta f}}. \quad (76)$$

Thus, provided we use a signal amplitude $\|\psi\|$ much greater than $\sqrt{N\sigma^2 \Delta f}$, we will have a nearly pure state. In practice, signal levels on the order of 10 V or so would be used, limiting the number of qubits, again, to about 40. Larger amplitudes, longer periods, and lower temperatures will all yield states of greater purity.

### 8.2. Computational complexity

We consider now the difficulty of constructing a quantum emulation device of the sort we have described. Bandwidth restrictions already place a limit on the number of qubits, so the question we address here is the physical resources required to construct a finite system. The device consists fundamentally of three basic analog electronic components: four-quadrant multipliers, operational amplifiers (serving as both inverters and adders), and bandpass filters.

The construction of two-qubit gates will dominate the resource requirements. General operations on qubits $i$ and $j$, where $i, j \in \{0, \ldots, n-1\}$ and $i \neq j$, require $n(n-1)$ distinct two-qubit gates. Each such gate utilizes a fixed number of multipliers, adders, and inverters, the total number of which scales quadratically with $n$. In addition, each of these gates utilizes a two-qubit projection operator, each of which requires two identical bandpass filters of the form $\mathrm{BPF}(n, i)$ as well as two pairs of identical bandpass filters of the form $\mathrm{BPF}(n-1, j)$. Each of the two $\mathrm{BPF}(n, i)$ filters must select $2^n/4$ distinct (positive) frequencies, while each of the four $\mathrm{BPF}(n-1, j)$ filters selects $2^{n-1}/4$ frequencies.

To construct an arbitrary bandpass filter over $2^n/4$ distinct frequencies, one could, for example, simply use a bank of $2^n/4$ tuned LRC circuits. Alternatively, as we have shown in section 3.3, a singe analog convolver, such as an elastic SAW device or optical CCD convolver, may be used in conjuction with an $(n-1)$-qubit masking signal to produce the desired bandpass filter. To achieve the necessary bandwidth, however, an analog convolver would require material structures within the device, such as the 'fingers' of an elastic SAW device, that scale with the required bandwidth. Either way, a density of structure is required that scales exponentially with $n$.

Current semiconductor technology is capable of producing a single integrated circuit device with over a billion transistors. Such technology could be used to produce a comb-like bandpass filter with a comparable number of narrowband frequencies, corresponding in this case to about 30 qubits. To achieve 40 qubits, the upper bound for a 1 THz bandwidth signal, a factor of 1000 increase in transistor density may be needed. Current trends in Moore's law predict that this may be achieved in the next 20–30 years, although advances in manufactoring may accelerate this development [24].

So, why not just simulate it digitally? A quantum state of $n$ qubits could be represented on a classical digital computer by storing each of the $2^n$ complex components in memory as, say, a pair of double precision floating point numbers. In this respect, the hardware requirements are comparable to that of a quantum emulation device, as described above. However, to perform gate operations one would also need to perform explicit matrix operations, which would be performed serially and scale in time as $4^n$. By contrast, a quantum emulation of a gate operation would be performed simultaneously over the entire Hilbert space. The inherent parallelism of quantum emulation is, then, one of the key characteristics that distinguishes it from mere simulation.

This is not to say that a quantum emulation is necessarily more efficient than a Turing-equivalent machine, which would be a violation of the Church–Turing thesis. A search problem that requires evaluating a function over $2^n$ possible inputs can be done serially, in exponential time, or in parallel, over exponentially many processors, using a classical digital computer. In either way, an exponential scaling of resources, whether in time or space, is required. In a similar manner, a quantum emulation device with an octave spacing of qubit frequencies would be constrained by an exponential scaling of required bandwidth. What we can say is that, aside from the limits on scale, a classical emulation of a quantum computer is capable of exploiting the same quantum phenomena as that of a true quantum system for solving computational problems.

## 9. Conclusion

In this paper, we have described a classical emulation of a quantum system suitable for quantum computing. In the proposed QMT scheme, the quantum state is identified with a complex basebanded time-domain signal. This signal may be represented as two separate real signals (representing the real and imaginary parts) or used to modulate a high-frequency carrier to yield a single real signal. An $n$-qubit state is represented by a linear superposition over a tensor product basis of functions consisting of complex exponentials of distinct, octavely spaced frequencies. The $2^n$ complex coefficients defining the quantum state in the computational basis are therefore encoded in the complex amplitudes of the $2^n$ component frequencies resulting from the sums and differences of the $n$ qubit frequencies. Time averages of signal products are used to define an inner product function, so this representation provides a complete Hilbert space description.

The operation of unitary gates is performed within this scheme via projection operators, which are themselves represented in terms of frequency-selective bandpass filters. One such filter may be constructed to address each particular qubit, pair of qubits, or, if needed, group of qubits. The complexity of these filters increases exponentially with the number of qubits, but their required passbands are easily defined and depend only on the qubit to be addressed. In addition, a convolution method was described that may facilitate the implementation of these filters. Measurement gates similarly rely upon projection operators to address specific qubits and use additional random variables to reproduce quantum statistics.

Because the QMT representation uses frequencies to encode the state, the number of qubits would be limited by bandwidth. Using an electronic voltage as a physical representation, we find that frequencies in the practical range of 0.1 Hz to 100 GHz would allow for about $n = 40$ entangled qubits to be realized. This number would be sufficient for useful quantum simulations or other low-qubit applications. Considerations of thermal noise, which has the effect of a depolarizing channel, yield similar limits. If needed, quantum error correction protocols could be applied to increase state fidelity, albeit at the cost of additional bandwidth requirements. Since the quantum state is explicitly encoded in the signal spectrum, security applications such as quantum key distribution would not seem to be amenable to this approach, although they may be emulated.

In summary, we have described a universal quantum computer built of classical components, fundamentally an analog computer that emulates the behavior of a true quantum system. A key advantage of this approach is the ease of construction, using standard integrated circuit technology, and robustness to decoherence. Quantum phenomena such as coherent superpositions and entanglement are faithfully emulated and leveraged to achieve an inherent parallelism. Thus, for any quantum algorithm for which the number of gates scales polynomially with the number of qubits, the required computation time will scale similarly. The practical limitation of this approach is in the number of qubits that may be represented. It remains to be shown that this approach can be realized in a fault-tolerant manner, using existing quantum error correction techniques, and that it ultimately has utility and merit over current approaches to quantum computing—or standard digital computers, for that matter.

## Acknowledgments

The authors would like to thank Mr Jitae Alex Kim for developing much of the Matlab code used to implement various quantum computing algorithms. We would also like to thank Dr Hans M Mark, Dr E C George Sudarshan, Dr Matthew E Thrasher, and Dr James E Troupe for many helpful and thought-provoking discussions. This work was supported by an Internal Research and Development grant from Applied Research Laboratories, The University of Texas at Austin, and is the subject of a provisional patent. Additional support was provided by the Office of Naval Research under Grant No. N00014-14-1-0323.